\documentclass[a4paper,fleqn,usenatbib,useAMS]{mnras}

\usepackage{color}
\usepackage{graphicx}	
\usepackage{amsmath}	
\usepackage{amssymb}	
\usepackage{multicol}        
\usepackage{bm}		
\usepackage{pdflscape}	

\usepackage{subfigure}
\newcommand{\sor}{GRS~1724--308 }
\newcommand{\aql}{Aql~X--1 }
\newcommand{\sorf}{GRS~1724--308}
\newcommand{\intg}{\emph{INTEGRAL}}
\newcommand{\xte}{\emph{RXTE}}

\usepackage[T1]{fontenc}
\usepackage{ae,aecompl}
\usepackage{caption}

\title[The failed state transition of GRS~1724--308]{The failed state transition of the ATOLL source \sor}
\author[A. Tarana]{A. Tarana$^{1,}$,  F. Capitanio$^{2}$\thanks{E-mail:fiamma.capitanio@iaps.inaf.it}, M. Cocchi$^{2}$\\
$^{1}$ ISS P. Baffi, via Lorenzo Bezzi 53, 00054 Fiumicino (Rome), Italy\\
$^{2}$IAPS/INAF, via del Fosso del Cavaliere 100, 00133 Rome, Italy\\}

\begin{document}

\date{Accepted anno mes giorno. Received ; in original form }

\pagerange{\pageref{firstpage}--\pageref{lastpage}} \pubyear{2018}

\maketitle

\label{firstpage}

\begin{abstract}

{The 2004-2012 X-ray time history of the NS LMXB \sor shows, along with the episodic brightenings associated
to the low-high state transitions typical of the ATOLL sources, a peculiar, long lasting ($\sim 300$ d) flaring event, observed in 2008.
This rare episode, characterised by a high-flux hard state, has never been observed before for \sor, and in any case  is not common among ATOLL sources.
We discuss here different hypotheses on the origin of this peculiar event that displayed the spectral signatures of a failed transition,  similar in shape and duration to those rarely observed in Black Hole binaries. We also suggest the possibility that the atypical flare occurred in coincidence with a new rising phase of the 12-years super-orbital modulation that has been previously reported by other authors.\\
The analysed data also confirm for \sor the already reported orbital period of $\sim 90$ d.}

\end{abstract}


\section{Introduction}
\label{intro}
\sorf, alias  4U~1722--30 {or H1724-307}, is a LMXB containing a weakly magnetized NS, revealed by the presence of type-1 X-ray bursts in its X-ray light curve \citep{swank}. {The bursts are typically Eddington limited \citep{cocchi} thus allowing to determine a distance of { about 9.5} kpc to the source \citep{kuulk}.} 
\sor is located in the globular cluster Terzan 2 \citep{grindlay80} and even if it is {known} since 70s that the cluster {hosts} a bright X-ray source, the \sor  position has been set {only {{in}} 2002~\citep{Revn}} by \emph{Chandra}, at confidence level above 4$\sigma$ within the radius containing half of the cluster mass. \\
\sor is one of the first NS systems known to have hard X-ray emission (E $>$35 keV) as revealed by SIGMA in 1991 with a power law emission extending above 100 keV and a photon index $\Gamma$ $\sim$1.65  \citep{barret91}.
{Its spectral/timing properties are typical of the ATOLL sources \citep{olive, altamirano}; the source spends most of the time in a low hard state (LHS) with occasional flares signalling a transition to the high soft state (HSS).}
The first detailed broad band study performed with \emph{BeppoSAX} and \emph{RXTE} showed a {LHS} Comptonized spectrum extending up to 200 keV of temperature $\sim 30 $ keV, plus an additional soft {disk} component of $\sim 1.5$ keV \citep{guainazzi}.

 \subsection[]{The long term monitoring of \sor\/}
 \label{222}
\begin{figure*}
\centering 
   \includegraphics[width=60mm,angle=90]{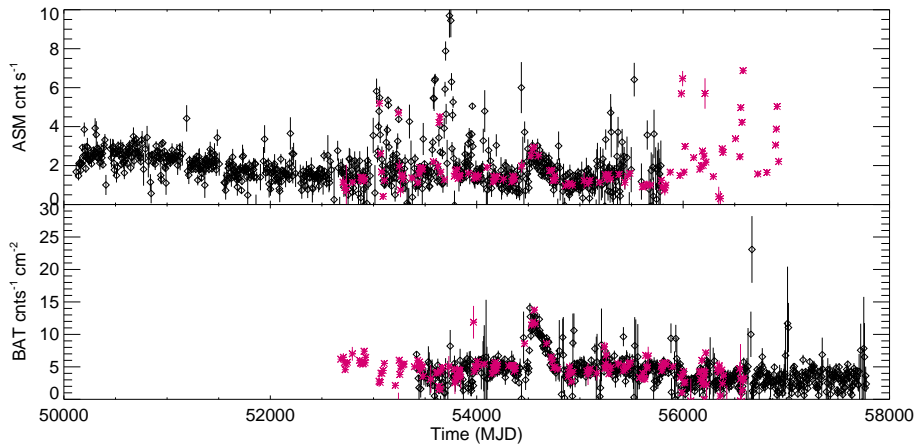}
  \caption{ Top panel: ASM/\xte\/ 2-12keV 10 days bin light curve of \sor\/. The pink asterisks represent the  \intg\//JemX 2-12 keV 10 days bin light curve rescaled in arbitrary units. Bottom panel:  \emph{Swift}/BAT 15-50 keV 10 days bin light curve of of \sorf. The pink asterisks represent the  \intg\//IBIS 20-100 keV 10 days bin light curve rescaled in arbitrary units. }
  \label{clucetot}
\end{figure*}

 Our study reports on the monitoring of \sor over 12 years. This period of time covers all the RXTE/ASM operational life plus an year of data collected only by INTEGRAL and SWIFT satellites {(see Fig.~\ref{clucetot})}.
However, exploiting further literature informations, we reconstructed the source behaviour over about 40 years that we resume herewith:

\begin{itemize}

\item {\bf MJD 42000--48000:}\\
Observations with Uhuru, OSO-8, Einstein and EXOSAT showed the source with a flux up to 30 keV below 30 mCrab with no particular flare events \citep{Forman, grindlay80, swank}.\\

\item {\bf MJD 48000--52000:}\\
 \citet{emel2002}, reported on a \xte\//PCA observation that showed  a flux increase throughout a long rise time of about 5 years, that reached a peak around the end 1996/early 1997 (MJD 50300-50500) and then showed a long decay phase, again of about 5 years. 

This modulation of the flux was tentatively associated to an intrinsic accretion rate variation from the donor star, or the passage of a third body, or even to a gravitational microlensing effect.\\

\item {\bf MJD 50000--53550:}\\
As metioned before, \citet{guainazzi}  reported on X-ray broadband observations performed with \emph{BeppoSAX} satellite in August 1996.The data were collected before and after a { type} I X-ray burst and revealed a spectrum that extends up to 150 keV, typical of a LHS.
This behaviour appears to be very similar to that of an observation performed by ASCA and RXTE in 1995 and reported by \citet{barret99}. \\
The \xte/PCA observations  published by \citet{altamirano} show that the source spent a long period (from MJD 51214 to 52945)  without any particular variation with a flux of about 25 mCrab in the 2-20 keV band.{ This period is partly overlapping and in agreement with the decay pha(for  significant HIDs).
se of the long term peak of the observations reported by Emelyanov (see above)}. After MJD 53000, the source changed its behaviour, and  showed different flares with peak fluxes up to 80 mCrab, { until the end of the observations.} A Lomb-Scargle periodogram in the PCA 2-20 keV band data showed a periodicity of about 90.55 days. This periodicity could be related to the orbital period of the binary system.\\

\item {\bf MJD 50000--55000:}\\
Kotze  \citep{kot} studied the long term X-ray variability in a sample of LMXBs, including \sor\/, to search for super-orbital periods using the ASM data from 1996-2009. Even though the period of observation is not enough to include more than one cycle, the source seems to have a super orbital modulation of 12$\pm$3 years. Such a long-term modulation has been supposed  to be produced by variation of accretion rate due to long magnetic-activity cycle of the donor star, a mechanism that has been proposed firstly in the Cataclysmic Variable objects \citep{CVref1, CVref2}.

\end{itemize}

\begin{figure}
\centering 
 \includegraphics[width=60mm,angle=90]{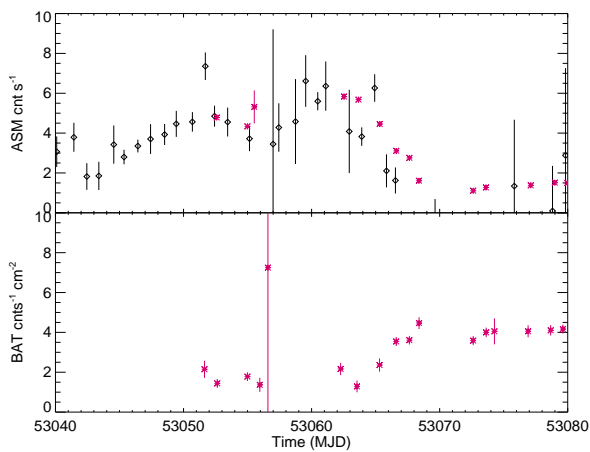}
  \caption{Example of  soft and hard flux anti-correlation during  a typical state transition flare of \sor\/. Top panel shows the  1 day averaged 2-12 keV ASM light curve super imposed with the Jem-X light curve at the same energy range. The bottom panel shows the 20-30 keV IBIS light curve. The INTEGRAL curves have a 1 day time bin. This flare occurred in 2004, before the launch of the Swift satellite,  thus the BAT data are not available.}
  \label{zoomcluce}
\end{figure}

\section{Data analysis and observation sets}
\label{data}
In order to follow the flux evolution of the source we  used the near-continuous coverage of the daily light curves from the  \xte\//All-Sky Monitor (ASM) \citep{levine} and \emph{Swift}/Burst Alert Telescope (BAT) \citep{krimBAT} monitors. The daily light curve of the soft X-ray monitor (2-12 keV) ASM are provided by the ASM/RXTE teams at MIT and at the RXTE SOF and GOF at NASA's GSFC and can be downloaded from the public web site\footnote[1]{ http://xte.mit.edu/ASM$\_$lc.html}. The daily light curve of the \emph{Swift} hard X-ray monitor BAT are available from the public web site\footnote[2]{ http://swift.gsfc.nasa.gov/results/transients/} and provided by the Swift/BAT team. 
{ The BAT/ASM {hardness} ratio has been calculated using the 1-day averaged light curves of both instruments expressed in Crab units and selecting the simultaneous observations. We used the conversion values reported in ~\citet{Yu}: 1Crab=75 cnt/s for ASM; 1 Crab=0.23 cnt/s/cm$^2$ for BAT.}

The IBIS and JEM-X light curves were collected using the HEAVENS tools provided by the INTEGRAL Science Data Centre web site~\citep{Walter}\footnote[3]{http://www.isdc.unige.ch/integral/heavens}.

The \intg\/ data, used for the spectral analysis, were reduced  using OSA 10.2 \citep{gol,winkler}, and include the X-ray monitor JEM-X \citep{lund} and ISGRI (15 keV--1 MeV) \citep{ubertini}, the low-energy detector of the $\gamma$-ray telescope IBIS \citep{lebrun}.
For the spectral analysis, XSPEC software v.12.7.1 \citep{xspec} has been used. Within XSPEC, a systematic error of 0.02 has been added to the spectral data, the instrument constant was fixed to 1 for IBIS and kept free for JEM-X.

The periodicity study performed on these data sets, has been performed by the XRONOS package v. 5.21.


\begin{figure}
\centering
    \includegraphics[width=50mm, angle=90]{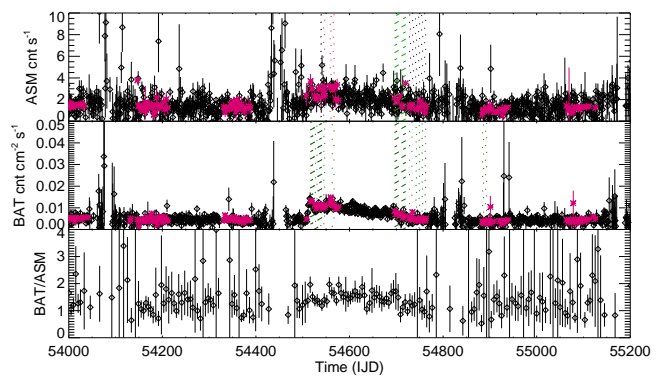}
   \caption{{ First two panels:} ASM and BAT light curves { as in} Fig~\ref{clucetot}, but zoomed around the { peculiar} 2008 { long} flare { (bin time 1 day)}. The green zones of the plots represent the INTEGRAL { observing periods throughout the flare} (during the periods represented with dashed zones the source has been observed only by IBIS, { while} during the dotted periods the source has been observed by both IBIS and JEMX). { Bottom panel: ratio between BAT and ASM fluxes (bin time 5 day)}}
  \label{zoom2008}
\end{figure}

\section[]{Time evolution of the source}
\label{time}
{ The first two panels of Fig.~\ref{clucetot} show the 2-12 keV RXTE/ASM and 15-50 keV Swift/BAT light curves, respectively (black diamonds), The pink asterisks represent the JEM-X (top panel) and the IBIS light curves (bottom panel) extracted in the same energy range of ASM and BAT, respectively, and rescaled in order to obtain similar  average counting rates}.

\subsection{The state transition flares}
\label{flares}
Fig.~\ref{clucetot} shows \sor is clearly variable, presenting several {flare-like episodes with typical durations of $\sim 30$ days}  most likely related to the spectral state transitions typical of the ATOLL sources.
These flare-like episodes show a peak flux in the soft X-rays ($<$ 20 keV) usually characterised by a short-time rise and a slower decay. The soft flux increase is generally followed by a delayed increase in the harder band ($>$ 20 keV).
As an example, in Fig.~\ref{zoomcluce} we show the ASM and ISGRI light curves of a state transition flare-like event occurred on 2004, where the anti-correlation between the soft and hard flux is evident.

\subsection{The 2008 long peculiar flare-like episode}
\label{2008flux}
Fig.~\ref{clucetot} clearly shows that roughly between { MJD} 54500--54900, the light curves present an evident simultaneous increase of the flux in both the soft and the hard energy bands.

This episode (hereafter termed  \emph{2008 flare}) is displayed in a zoomed view in { the first two panels of} Fig.~\ref{zoom2008}. The 2008 flare has {\emph{profile}} and  {\emph{duration}} very different from the ones of the state transition flares. In fact, this smoothly varying event shows NO anti-correlation in the hard and soft energy bands and lasted for more than 200 days (54500 MJD - 54725 MJD), i.e.  $\sim 7$ times the typical duration ($\sim 30$  days) of the state transition flares. 
The intensity ``bump'' is clearly evident in the BAT 15-50 keV {lightcurve}. During this peculiar flare the average hard flux (>20 keV)  reached a peak value never observed before in the \sor history. 
{ Moreover, {there is no indication of significant hard colour variability during the flare}
(see Fig.~\ref{zoom2008}, 3$^{rd}$ panel).}

\subsection{Periodicity studies}
\label{period}
  If  we {assume} that the first maximum of the super orbital modulation reported by~\citet{kot}  {coincides} with the peak flux {in late 1996 -- early 1997} reported  by \citet{emel2002} ({i.e}. the authors reported on the same super orbital {cycle}), the {2008 flare occurs} at the beginning {of the next super orbital cycle} i.e., {within the errors, at the new minimum}.
{Of course this statement needs to be verified on a much longer time basis including more than a single} cycle of long term modulation. \\

{Our short term} periodicity studies confirm the $\sim$90 days period {firstly reported } by~\citet{altamirano} in the PCA data.
{The 90 d period} is present {both in our ASM and BAT data sets}.
{For this periodicity,} the folded light curves of both ASM and BAT are shown in Figure~\ref{period}. 
This modulation i{s most likely related} to the orbital period of the binary system {\citep{altamirano}} and not to the super orbital modulation proposed by ~\citet{emel2002}.

\begin{figure}
\centering
\includegraphics[width=50mm,angle=-90]{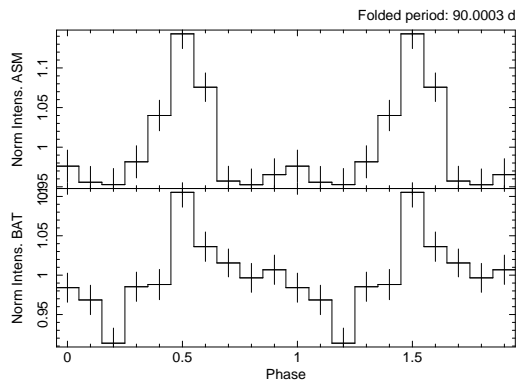}
\caption{Top panel: ASM folded light curve of \sorf. The period corresponds to 90 days. Bottom panel: BAT folded light curve { for the same period.} }
  \label{period}
\end{figure}

\section{Colour--Intensity diagrams}
\label{col}
We used the INTEGRAL data to construct hard colour { vs} intensity (or hardness--intensity) diagrams (HID). 
{JEM-X counts have been collected} in two energy ranges {(soft, 4-10 keV, and hard, 10-20 keV), along with simultaneous IBIS counts in the 20-30 keV range.}\footnote[4]{the ASM and BAT data are too noisy to produce 
good HIDs { but for the 2008 flare data.}}.
Two sets of HID have been constructed: hard/soft JEM-X ratio and ISGRI/soft JEM-X ratio, in function of the total count rate in the two bands.

\begin{figure*}
\centering

\subfigure
  {\includegraphics[width=50mm,angle=90]{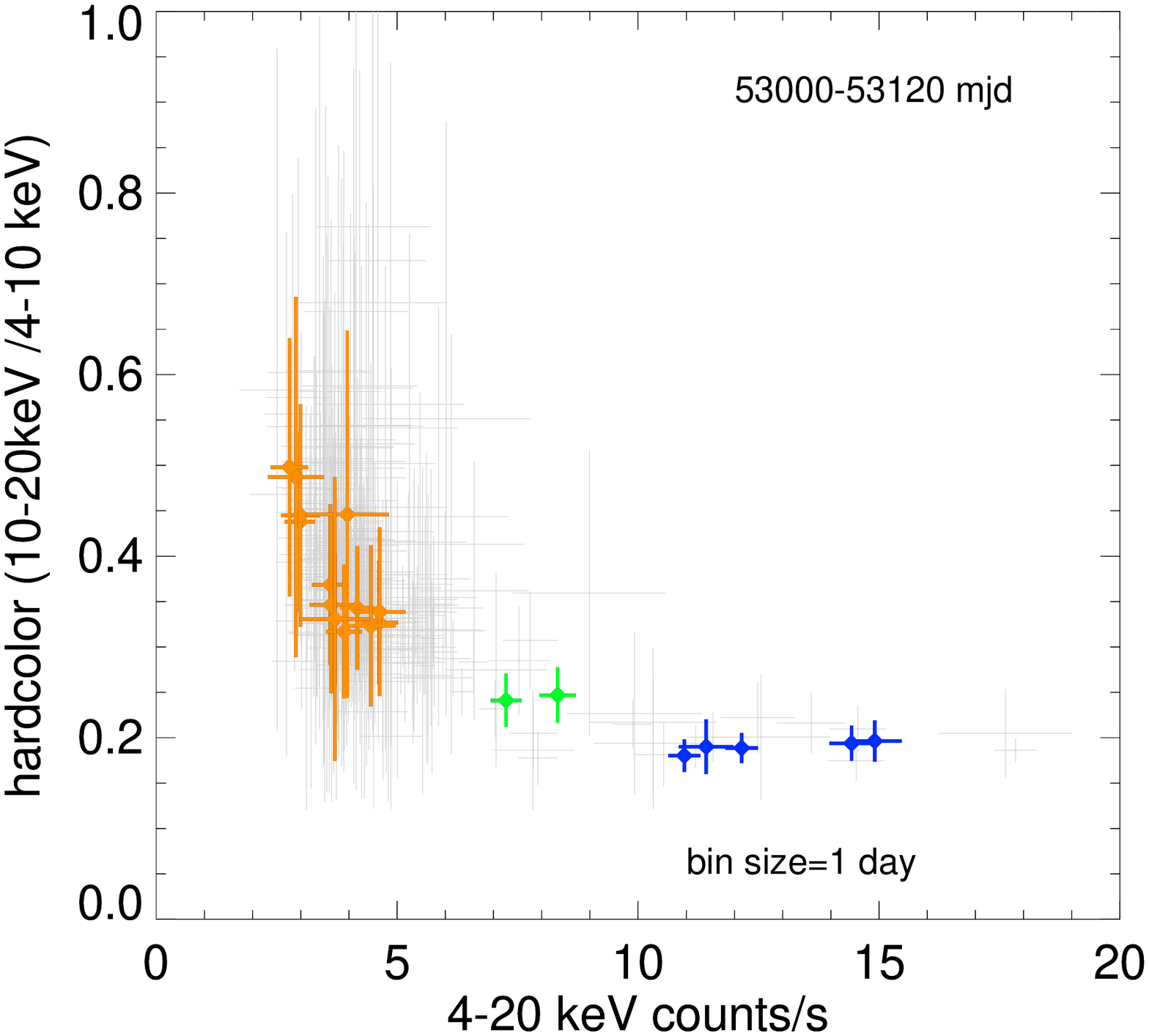}}%
{\includegraphics[width=50mm,angle=90]{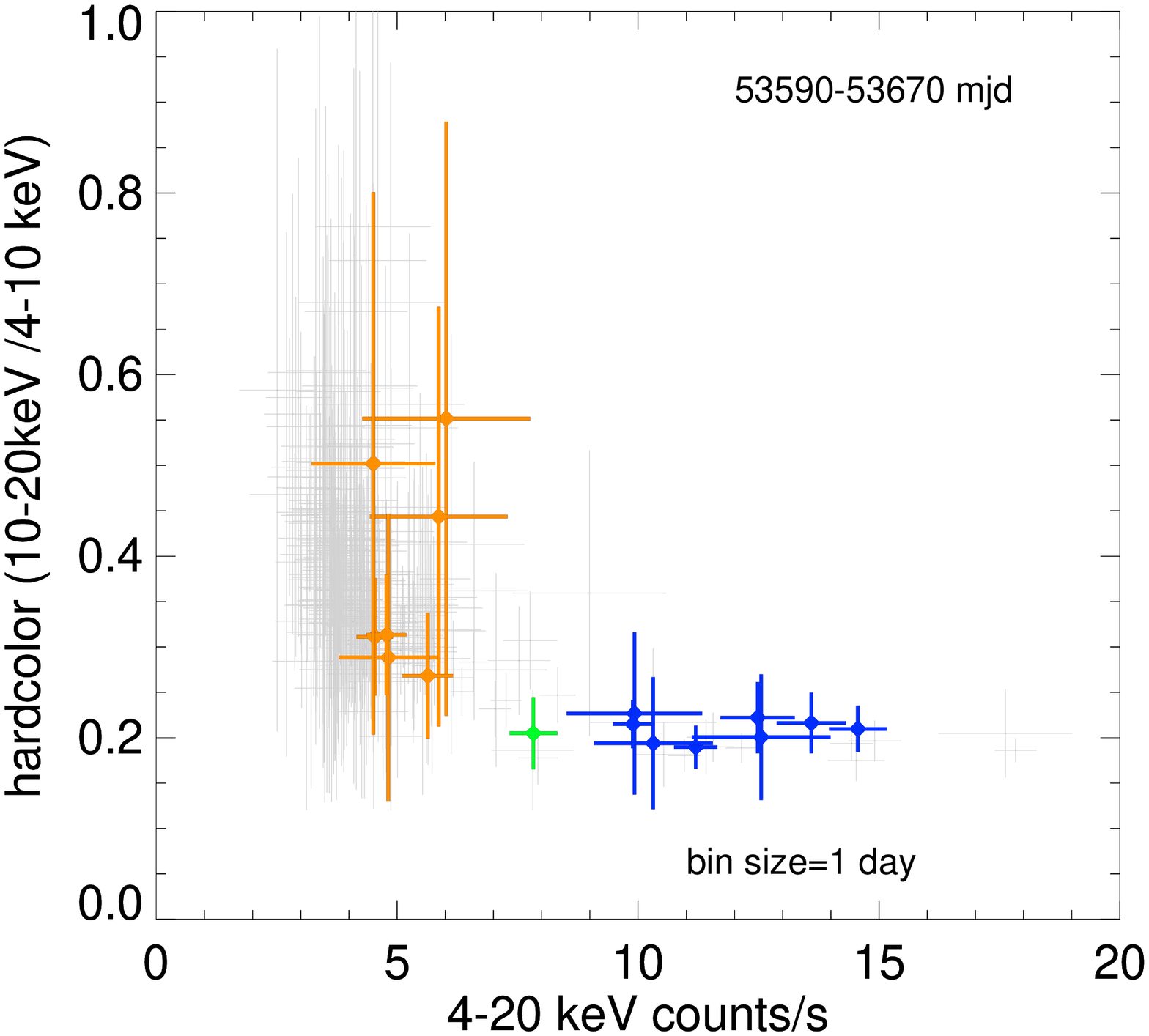}}%

\vspace{-6mm}
 \subfigure
  {\includegraphics[width=25mm,angle=90]{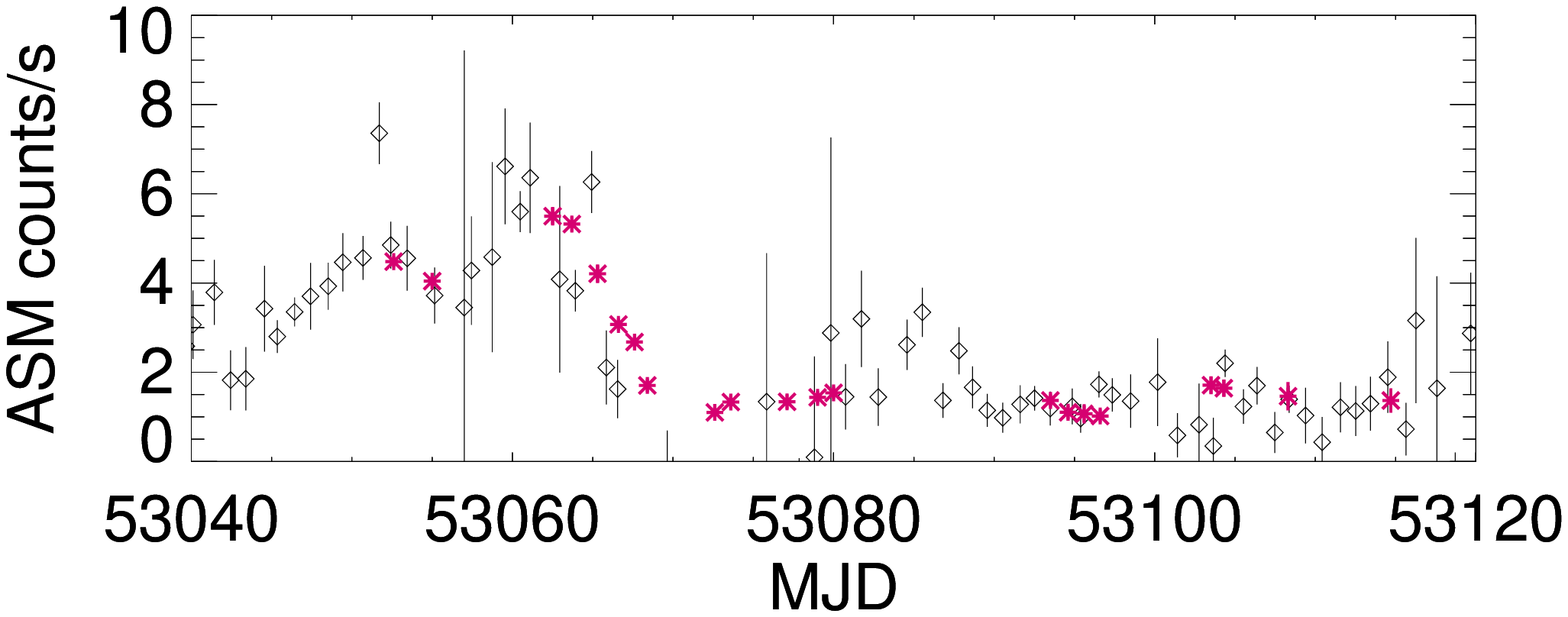}}%
{\includegraphics[width=25mm,angle=90]{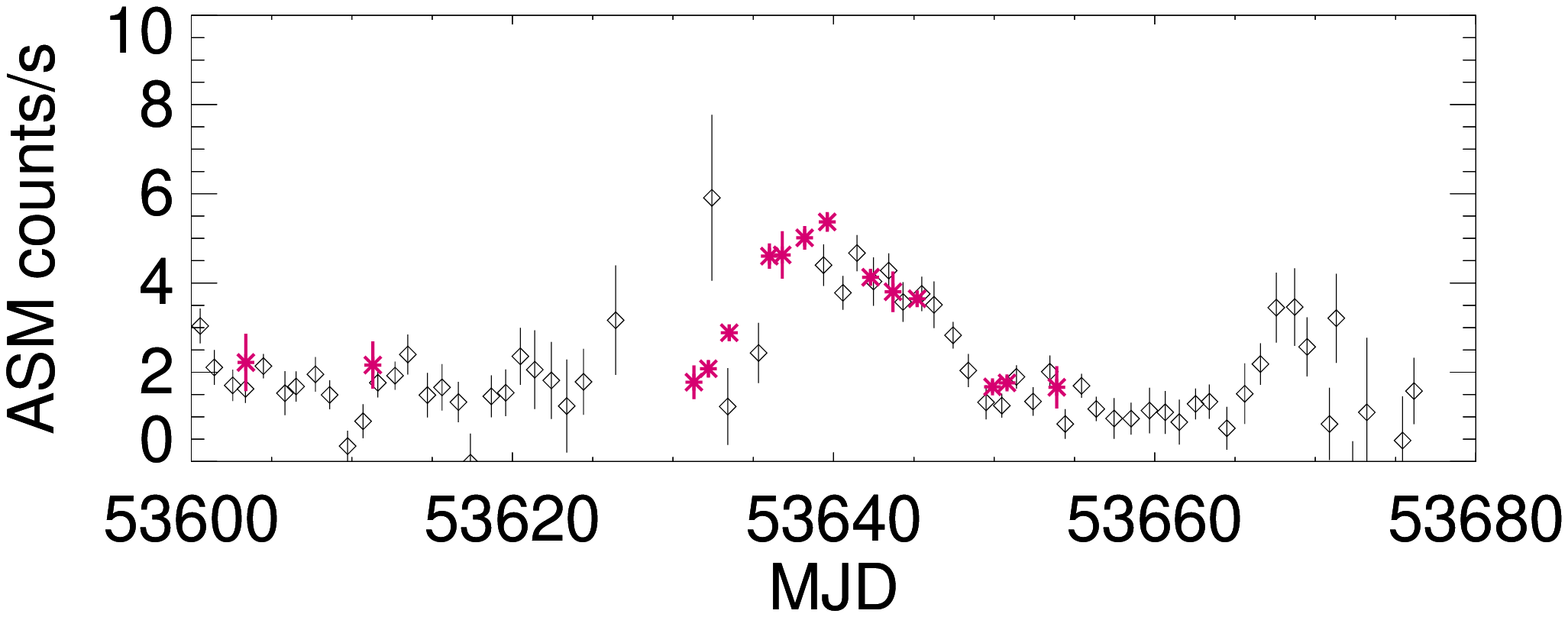}}%
   
 \subfigure
{\includegraphics[width=50mm,angle=90]{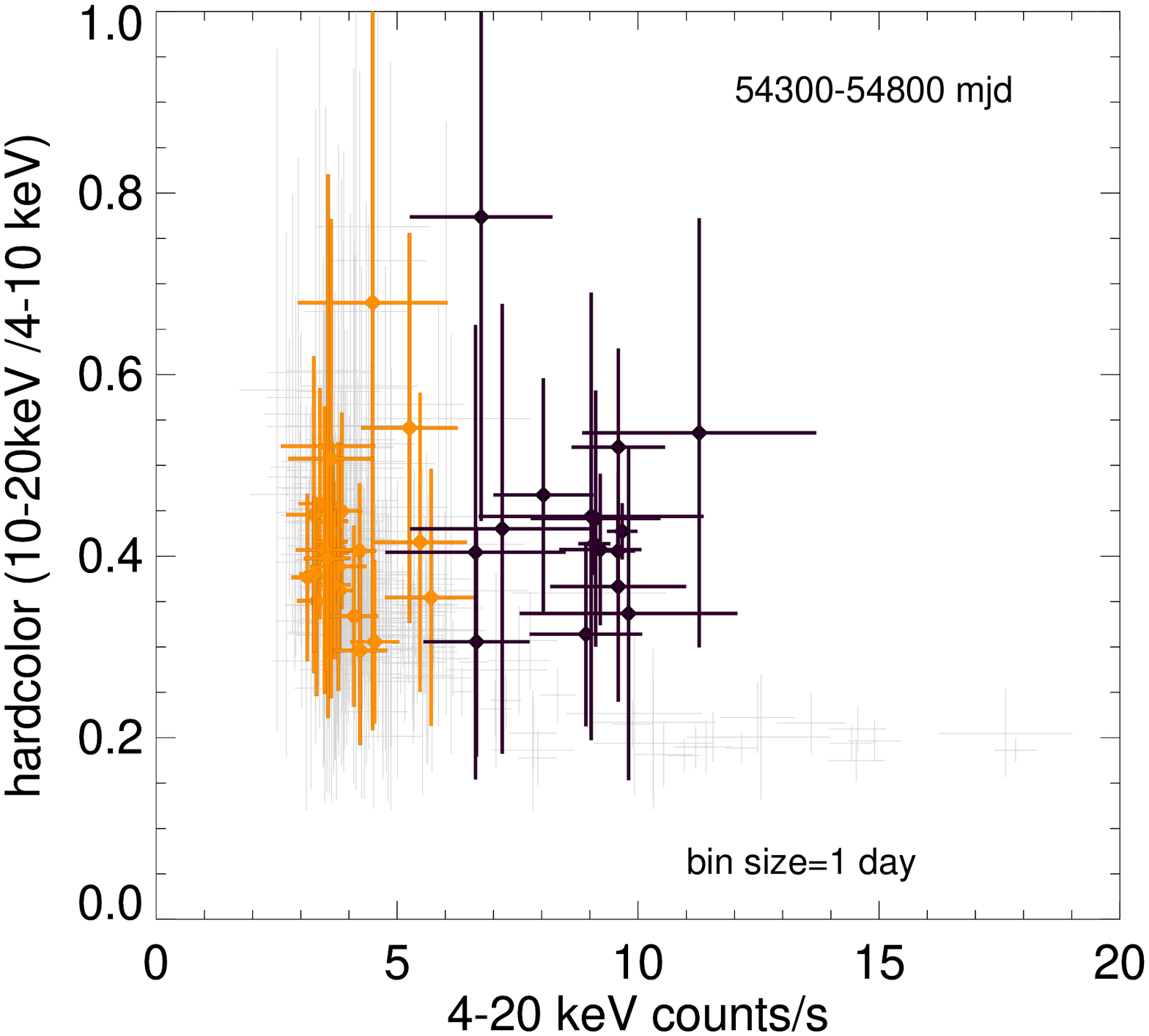}}%
\includegraphics[width=50mm,angle=90]{{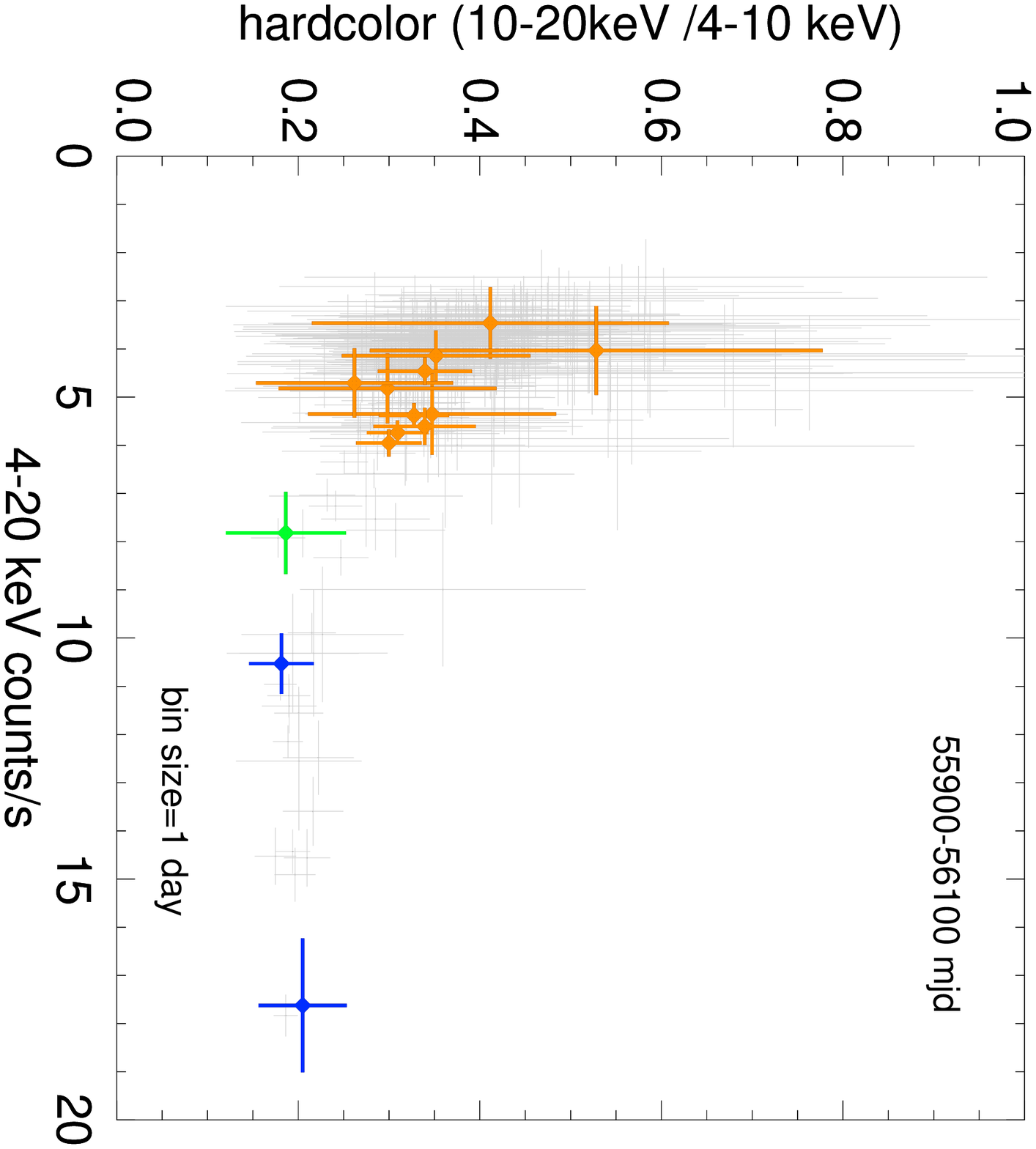}}%

\vspace{-6mm}
\subfigure
 {\includegraphics[width=25mm,angle=90]{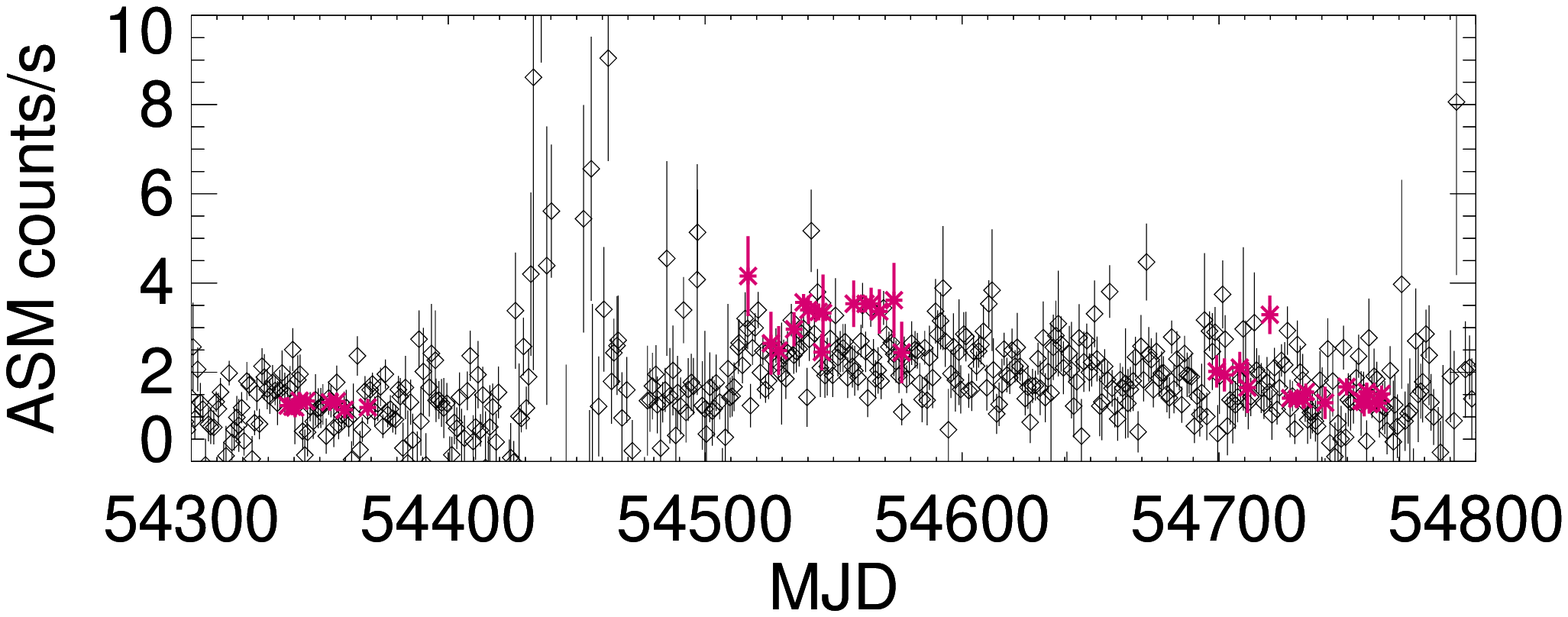}}%
{\includegraphics[width=25mm,angle=90]{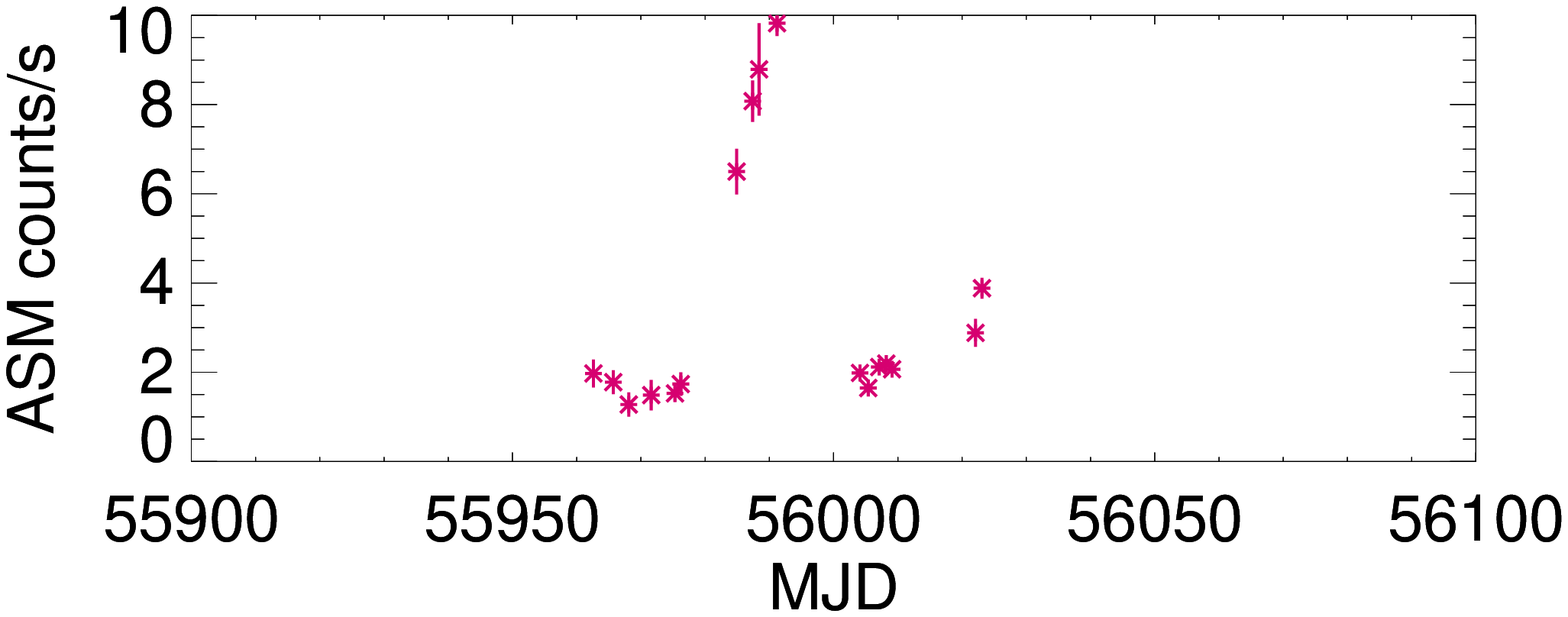}}%

  \caption{Jem-X color Intensity diagrams. The time bin of each point is equal to 1 day, the coloured points in each figure represent the evolution of  the state transition of a single  flare of the source. The points of the same colour represent the same spectral state (yellow for hard states, green for intermediate states and blue for soft and very soft states). Instead, the background grey data represent the whole dataset including the persistent emission.  The black points represent the peculiar state observed only during the 2008 flare (for clarity these points were omitted in the background grey data, in order to highlight the ATOLL source HCI track of \sor).}
  \label{hci_jmx}
\end{figure*}

{The JEM-X HIDs are displyed} in Fig.~\ref{hci_jmx}. Each HID diagram {refers} to a different flare (coloured points), whose points are over-plotted on the ones of the whole data set, including both flares and steady emission (background grey crosses).  The corresponding light curve is reported in the figure at the bottom of each HID.

For each diagram we have used different coloured symbols to group the data with similar spectral states. 
{ Spectra groups have been defined} imposing a fixed range of values of hard colour and intensity.
The orange points indicate the hard state data, the blue points the soft state data, the green {ones} the intermediate state data. 
The black points are {the ones of the 2008 peculiar flare}. 
As the HIDs show, the source spends most of its time in a hard state and during {an occasional} flare it typically shows a state transition.  During the various {``standard''} flares the source {tracks} similar curves in the diagram.  
{Conversely}, during the 2008 flare the source {exhibited} a {bright} hard state (black points) with a different position in the HID. This position was never reached {again} in the 12 years that are the subject of our study and this peculiarity was {previously unreported}.
 
\begin{figure}
\centering
    \includegraphics[width=75mm,angle=90]{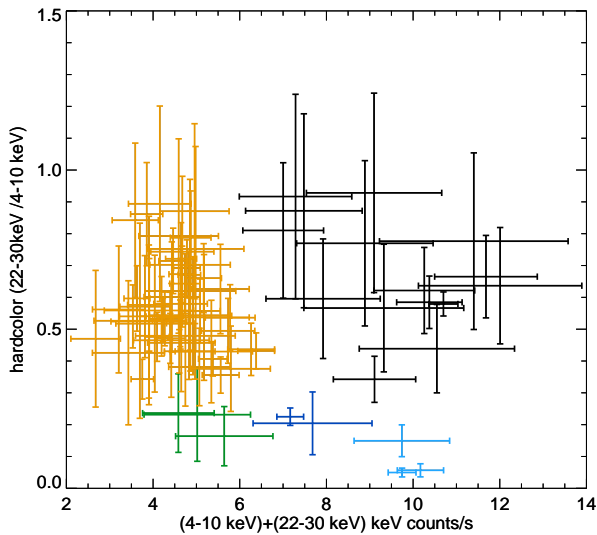}
   \caption{Hardness-Intensity diagram for IBIS/JEM-X simultaneous pointings. The time bin of each point is one day. the different colours  correspond to the same spectral states as used in Fig.~\ref{hci_jmx}: hard state (orange points), peculiar-hard state in 2008 (black points), intermediate state (green points) and soft state ( blue/light blue points). }
  \label{hci_ibis}
\end{figure}

Fig.~\ref{hci_ibis} {displays} the {HID} derived from simultaneous JEM-X and IBIS pointings throughout the monitoring period\footnote[5]{ Due to the INTEGRAL observation strategy and the different FOV of the two instruments, only part of the JEM-X and IBIS data are simultaneous.} 
{The colours of the points have the same meaning as in Fig. \ref{hci_jmx}}. 
The separation of the various {spectral} states, included the peculiar {bright} hard state of the 2008 flare, is evident also in this figure.

\begin{figure}
\centering
    \includegraphics[width=80mm,angle=90]{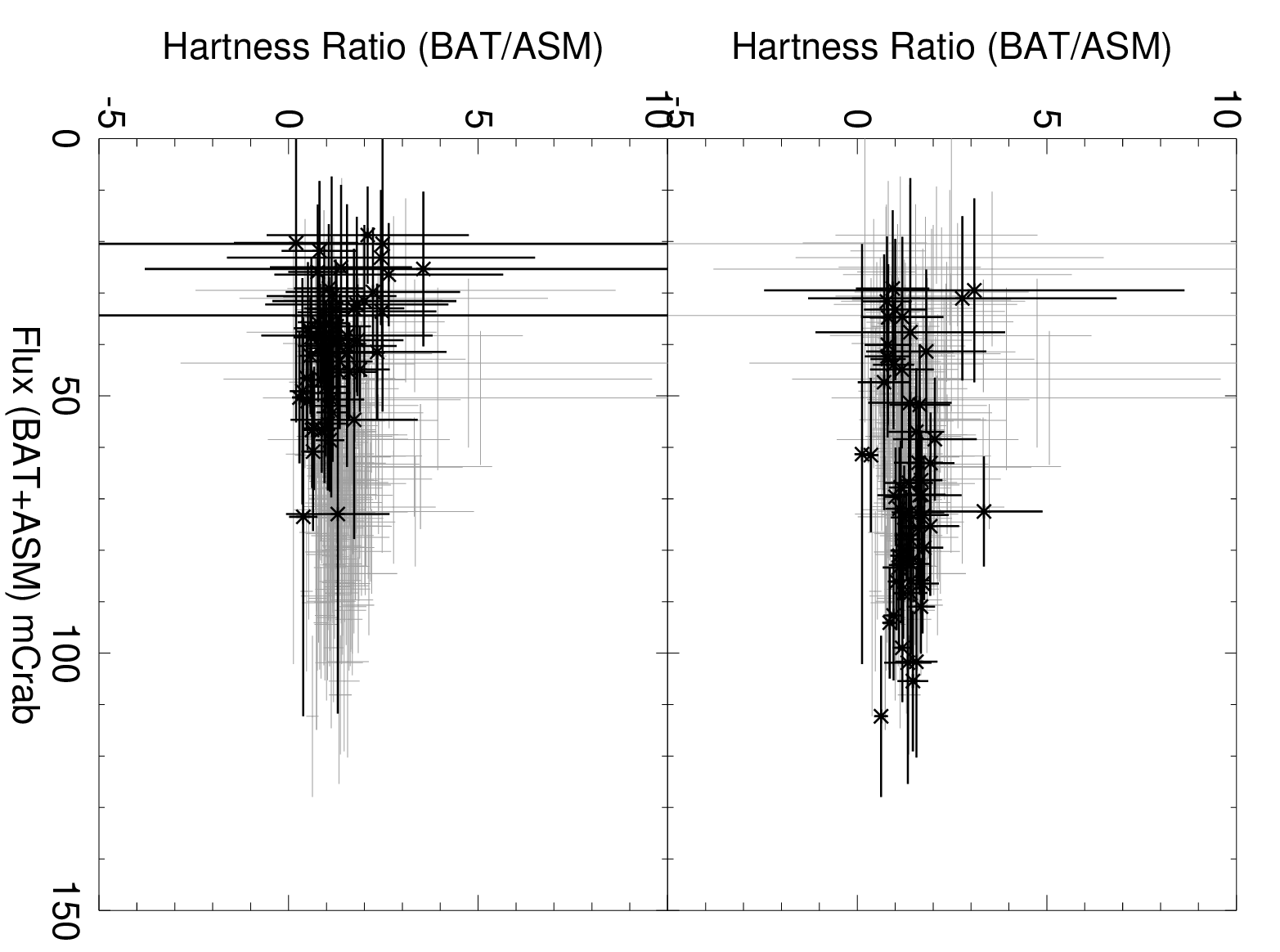}
   \caption{ First panel: BAT/ASM HID of the entire 2008 peculiar flare (grey points). The black points represent the 1$^{st}$ 60 days of the flare. 
   As in the top panel, for the last 60 days of the flare (in black). Time range is (MJD)=54489--54781, bin time is 1 day (see also Fig~\ref{zoom2008} for  light curve details}
  \label{BA_HID}
\end{figure}

\subsection{2008 flare: BAT/ASM HID}
{ {Due to observatory and time allocation constraints, INTEGRAL daily monitoring of \sor is not available and this could prevent the detection of possible short time-scale state transitions during the 2008 flare.
Nevertheless, the flare is bright enough, compared to the rest of the \sor time history, to allow the construction of a significant BAT/ASM HID of the event.
In particular, the first and last 60 days of the flare were investigated in comparison with its total HID evolution (Fig.~\ref{BA_HID}).
\\
The first panel of the figure shows a clear flux increase with constant hardness at the beginning of the flare.  In its last part, the flux decreases slowly (see also e.g. Fig~\ref{zoom2008}) and the flux drop is less evident in the HID of the last 60 days.
\\
{ The calculated averages of the hard colour are $1.50 \pm 0.58$ for the entire flare and 
$1.36 \pm 0.15$, $1.6\pm 2.9$ for the leading and trailing 60 d, respectively. 
Their consistency within the errors indicates there is no time evolution in the hard colour during the 2008 event, as already suggested by Fig~\ref{BA_HID}. Also The calculated $\chi^{2}$ for the 1 day ratios are compatible with no HR variability.}
}
}

\section{Spectroscopy of the source}
\label{spe}
\begin{figure*}
\centering
\includegraphics[width=80mm,angle=-0]{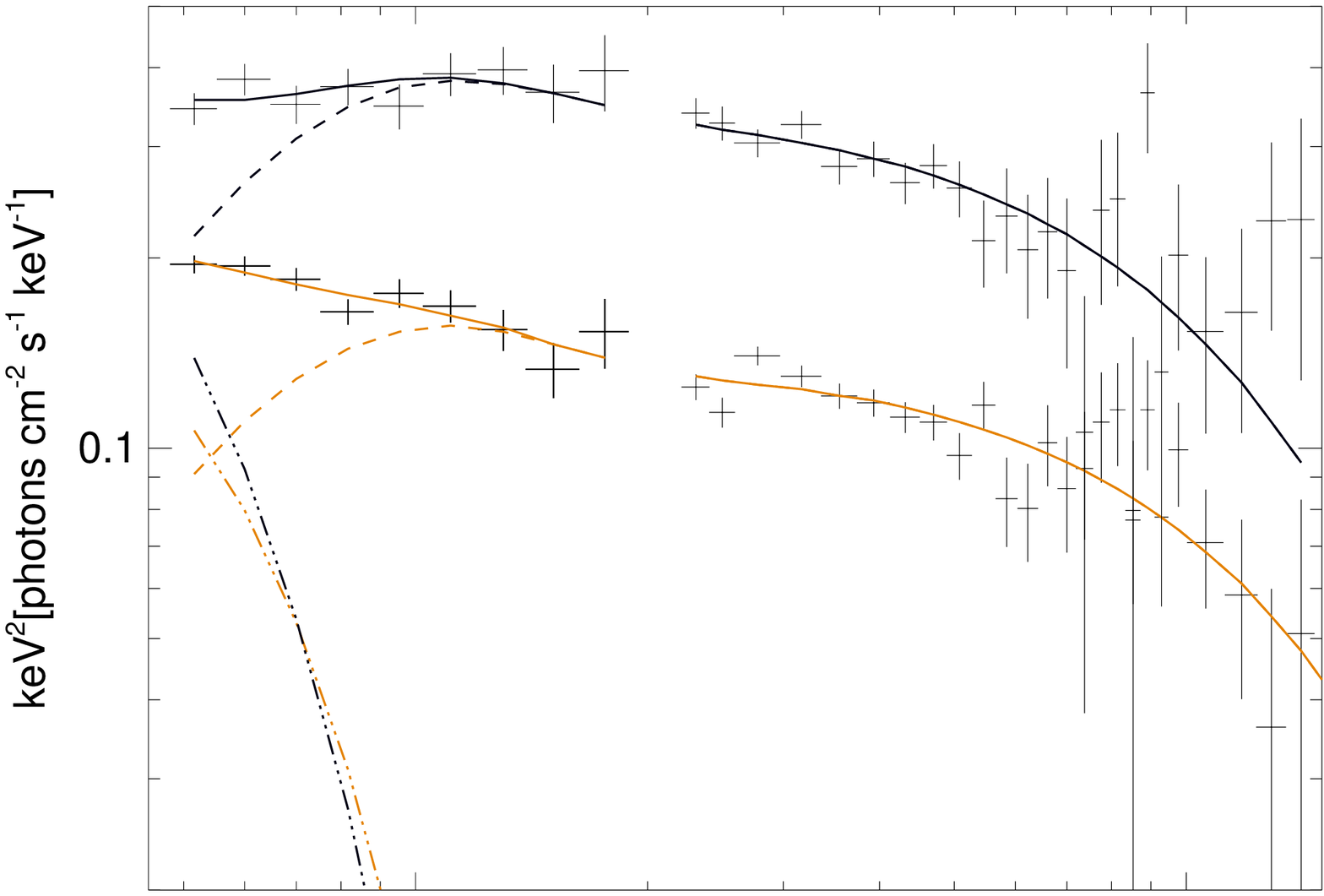}
\vspace{-9mm}

\includegraphics[width=80mm,angle=-0]{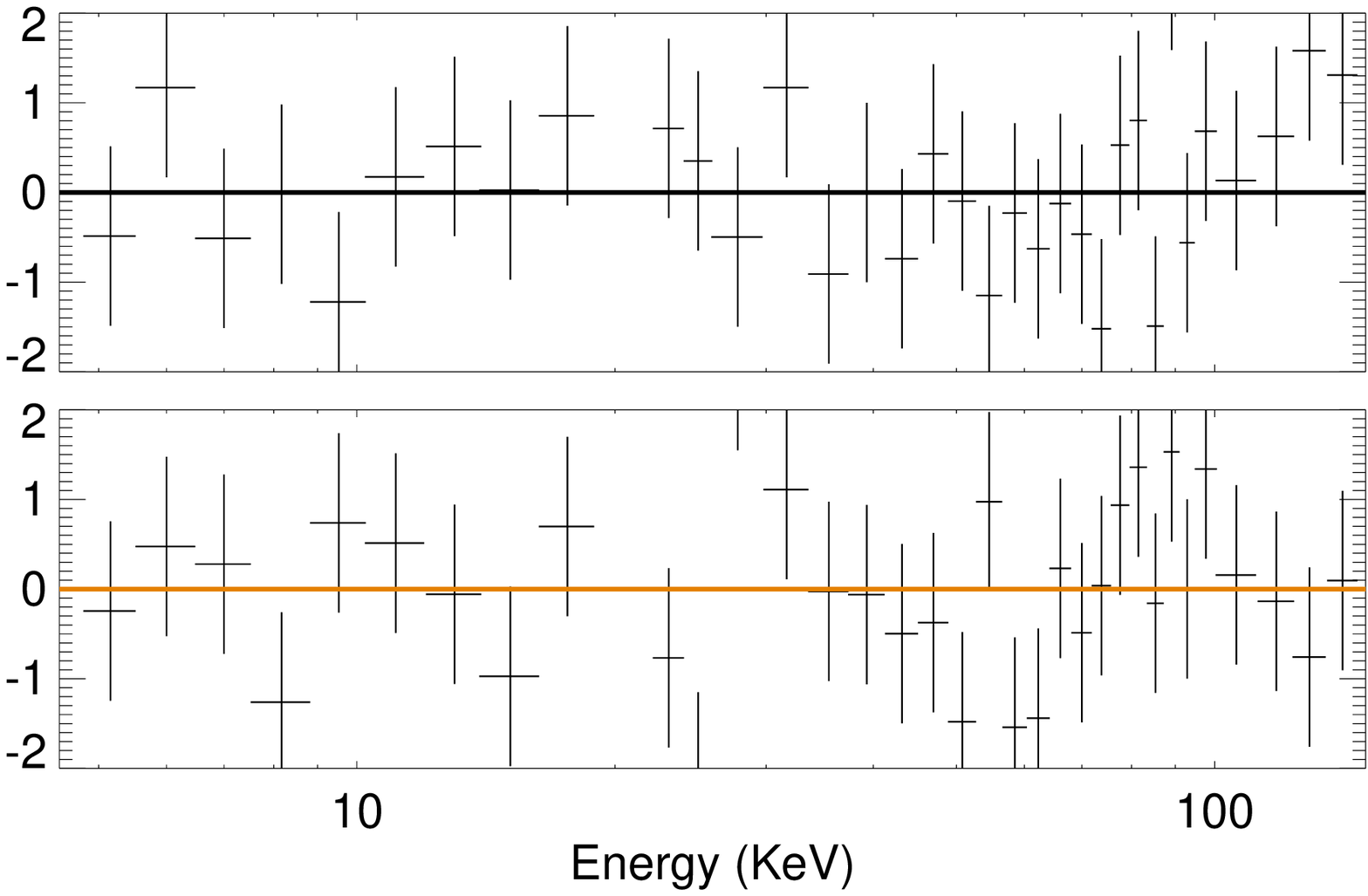}
\caption{Top panel: \sor\/ IBIS and JEM-X hard state unfolded spectra. The spectrum with orange lines corresponds to the orange points in Figure~\ref{hci_ibis} (regular \sor hard state.) The spectrum with black lines corresponds to the black points in Figure~\ref{hci_ibis}  (2008 peculiar hard state) . Middle and bottom panel: respective fit residuals in terms of sigmas with error bars of size one. }
  \label{spe_ora_bl}
\end{figure*}

We performed the analysis of the averaged simultaneous IBIS and JEM-X spectra constructed by collecting data of different flares with the same hardness characteristics, i.e. data with the same colour in the Jem-X/IBIS  hardness-intensity diagram (see Figure~\ref{hci_ibis}). 
Figure~\ref{spe_ora_bl} shows the spectra of the two hard states  of \sor { corresponding} to the orange and the black points in Figure~\ref{hci_ibis}. The model used for the fitting procedure is a disk black body component ~\citep{mitsuda} plus a comptonization of soft photon via relativistic electron plasma ~\citep{poutanen}, DISKBB+compPS in XSPEC. 
 CompPS is effective in modeling Comptonized emission from a hot, optically thin plasma region like that of the NS-LMXB in a LHS.
The plasma parameters commonly observed in LHS (kTe > $\sim 20$ keV, $\tau_{s}$ < $\sim 3$) are safely within the model's validity parameter space.
 Instead, when analytic models assuming a diffusion regime (e.g. the largely adopted CompTT, ~\citet{titarchuk}), are applied to low opacity plasma clouds,
the fit values obtained could be unreliable.

The { fitted parameters for} the two { LHS} { (see Table~\ref{tab_fit1722}} are consistent to each other within the errors and with the { ones} reported in literature for the { source LHS} \citep[{ e.g.}][]{guainazzi, barret99}.
The two spectra have instead very different flux{ : in} fact the 2008 peculiar { LHS} is { $\sim 50\%$} brighter than the standard { one}.
Although the rising of the 2008 flare is not covered by INTEGRAL observations, { the HR vs time of the final part of the flare  does not show variations within the errors, therefore}  we { presume} that the flare  substantially evolved at constant spectrum {  as already shown by the BAT/ASM flux ratio in Fig.~\ref{zoom2008} }.
\\
\\
{ As far as the standard flares HSS spectra are concerned, only Jem-X data have been used for spectral extraction}.
We obtained { good quality} spectra only for { the data sets} corresponding to blue and light blue points in Figure~\ref{hci_ibis}. 

We modelize the { soft} spectra with a disk black body component plus an { inverse Compton} component. In this case { the Comptonization model is CompTT} ~\citep{titarchuk}.
{ \texttt{CompTT} is effective for low temperature, highly opaque electron plasmas as the HSS ones}. Unfortunately the disk black body component could not be constrained because it is confused into the Comptonised one. In the hard state spectra, instead, the Comptonised component peaks at higher energy ($>$ 20 keV) and let the disk black body component clearly visible even with the Jem-X spectral capability.
Fitting the HSS data only with \texttt{CompTT}, we obtained spectral parameter values in agreement with those reported in literature for other { NS LMXB} \citep[see for example][and reference therein]{barret2000,ada,tarana08}.
Compared to LHS spectra, the electron plasma temperature decreases { to $\sim 3$ keV}, while the optical depth { $\tau$ exceeds the value of 5} { (see Tab~\ref{tab_fit1722soft}}).

\begin{table*}
\centering
\caption{Comparison of spectral fitting results for the JEM-X and IBIS hard state spectra of \sor. The model used for the fitting procedure consists in a disk black body component, Diskbb in XSPEC~\citep{mitsuda}, plus a comptonization model, COMPPS in XSPEC~\citep{poutanen}. A renormalising constant has been included in the spectral fit in order to take into account  calibration differences between the two instruments.  The colours corresponds to those used in the hardness-colour diagram (Fig.~\ref{hci_ibis}). All the errors and upper limits reported in the table correspond to the 90\% confidence level.}
\label{tab_fit1722}
\vspace{0.4cm}
\begin{tabular}{cccccccccc}
\hline
\hline
Spectrum &  Factor & Tin   & $Norm_{D}$ &   KTe &   tau-y & $Norm_{C} $& $\chi^{2}_{reduced}/d.o.f.$ & Flux$_{2-10 keV}$ & Flux$_{10-150 keV}$  \\
 --             &   --        & keV  &   ---               &   keV & ---       & ---                 & ---                                        &ergs~cm$^{-2}$~s$^{-1}$ &ergs~cm$^{-2}$~s$^{-1}$ \\
Orange & ${0.87}^{+0.18}_{-0.17}$ & ${1.20}^{+0.40}_{-0.34}$ &  ${12}^{+44}_{-9}$      & ${45}^{+15}_{-12}$     &   ${1.01}^{+0.52}_{-0.34}$ &  ${4.07}^{+0.66}_{-0.51}$ & 1.1/30 & 4.9e-10& 5.0e-10\\
Black  &   ${0.90}\pm{0.19}$         & ${1.03}^{+0.51}_{-0.43}$  & ${37}^{+226}_{-34}$   & ${40}^{+13}_{-10}$     &   ${1.13}^{+0.46}_{-0.35}$  &  ${10.2}^{+1.5}_{-1.2}$  & 1.0 /28   &9.2e-10 &12e-10\\

 \hline
\hline
\end{tabular}
\end{table*}

\begin{table*}
\centering
\caption{Spectral fitting results for the JEM-X soft state spectra of \sor. The model used for the fitting procedure consists in  a comptonization model, COMPTT in XSPEC~\citep{titarchuk}.  Because of the data energy interval starting from 3.5 keV, the seed photon temperature has been freeze to its best value (0.8 keV). With the addiction of a DISKBB component the fit parameters can not be constrained (see text for details). The data corresponds to the blue and light blue points in the hardness-colour diagram (Fig.~\ref{hci_ibis}). All the errors and upper limits reported in the table correspond to the 90\% confidence level.}
\label{tab_fit1722soft}
\vspace{0.4cm}
\begin{tabular}{ccccccc}
\hline
\hline
Spectrum  &      KTe                 &   tau-p                          & $Norm_{C} $                       & $\chi^{2}_{reduced}/d.o.f.$ & Flux$_{2-10 keV}$  & Flux$_{10-150 keV}$ \\
 --              &      keV                 & ---                                 & ---                                        & ---                                        &ergs~cm$^{-2}$~s$^{-1}$  & ergs~cm$^{-2}$~s$^{-1}$\\
 Light blue &${2.8}^{+0.6}_{-0.4}$    & ${5.5}\pm 1$  & ${0.08}\pm 0.01$  & 0.40/13                             & 8.4e-10& 2.3e-10\\
Blue          &${2.9}^{+1.4}_{-0.6}$ & ${5.5}^{+2.2}_{-1.9}$  &${0.06}^{+0.01}_{-0.02}$& 0.30/11 &6.5e-10 &1.9e-10\\
\hline
\hline
\end{tabular}
\end{table*}


\section{Discussion}
\label{disc}

{
The time history of \sor alternates long-lasting low luminosity periods of negligible activity to other phases, generally shorter, with strong activity characterized by bright flares and LHS to HSS transitions. This is a typical time behaviour of the persistent low-luminosity ATOLL sources. For this subclass, the state transitions to the HSS are relatively rare episodes as their canonical spectral state is a LHS of $L \sim 10^{36}~{\rm erg~s}^{-1}$.  Conversely, the bright persistent ATOLLs (e.g. most of the so-called GX sources) exhibit an HSS ($L > 10^{37}~{\rm erg~s}^{-1}$) as their canonical state, with no or rare transitions to the LHS.
The different behaviour supposedly reflects intrinsic differences in the binary systems, such as donor mass and activity, orbital parameters, star population, etc.
As an example, the persistent emitters located in a globular cluster are generally faint, LHS, ATOLLs.
Among the latter sample, \sor in Terzan 2 seemed to show no peculiarities.
\\
This work reports on the observation of an atypical, long lasting (>300~d) \textit{hard flare}, which is an unprecedented event in the source time history.
The 2008 flare, for its shape, time scale and coverage of an unusual bright-hard region in its HID, shares similarities with the failed state transitions (or failed outburts)}
observed { in some cases} in Black Hole binary systems~\citep{brocksopp,capitanio,ferrigno}.
{
The main difference is the persistence of the emission of \sor, as { almost} all the BH LMXB are transient sources.
\\
Among NS LMXB similar failed transitions are much less common. An interesting case is the one of \aql, a nearby, quasi-persistent (or quasi-transient) NS which exhibited a
}
failed state transition during an outburst \citep{rodriguez}.
However, { unlike the \sor one,} the failed outburst { of \aql} shows both duration and luminosity similar to { the standard outbursts of the source but in that case, according to the authors, the LHS to HSS state transition generally} observed during the outbursts { did} not occur. 
{ Moreover, even though \aql is} classified as an ATOLL source, \cite{rodriguez} { underline its} behaviour similar { to that of a number of} BH binaries, XTE 1550-564~{ in particular.
Conversely, as mentioned above, \sor has no peculiarities among its class of objects.
So the two episodes are likely intrinsically different in their nature.}
\\
{
For comparison sake, we investigated the critical luminosities for the atypical failed transitions of both \sor and \aql, here defined as the 
luminosities where the LHS to HSS transition is expected to occur.
Extrapolating the \aql unabsorbed cut-off power-law spectrum published by \cite{rodriguez}, a bolometric luminosity (for a 5 { kpc} distance) of $\sim 2\times 10^{37}~{\rm erg~s}^{-1}$ is obtained for the peak of the failed outburst. This value should equal, or be very close to, the critical one, as e.g. \cite{maccarone} report the 1999 HSS transition outbursts in the $1.6-2.2 \times 10^{37}~{\rm erg~s}^{-1}$ range.
For the case of \sor the bolometric extrapolation of the relevant spectra (see the sections above) lead to $L \sim 2.8\times 10^{37}~{\rm erg~s}^{-1}$ for the hard 2008 flare, very similar to the case of \aql.
The typical (years 2003--2005) state transition flares of \sor peak at { $L \sim 1.8\times 10^{38}~{\rm erg~s}^{-1}$ \citep{tarana08}, }which is a relatively higher value.
However, this value should be regarded only as an upper limit to the one of the critical luminosity for \sor.
One could reasonably deduce that the 2008 hard flare did not reach the critical luminosity for the LHS to HSS transition, so the flare should be regarded as the effect of an anomalous accretion event rather than a {\em classic} failed outburst episode. 
The 2008 perturbation, whose energy surplus has been likely almost completely dissipated} via coronal emission, { looks less powerful and acting on a much longer time scale than the standard ones.
\\
Such a rare, weak perturbation could be connected to}
the $\sim$12 yr super orbital modulation reported in Section~\ref{222}, { possibly related to a secular magnetic cycle of the donor star.
In fact} the peculiar 2008 flare { looks} in coincidence with the minimum { of the long term period, though} we do not have enough data to { establish a firm correlation}.
\\
{ Besides this possible long term coincidence, a rare variability episode of the donor, or the periastron of a third body in an eccentric orbit, or the transit of a nearby star could explain the anomalous 2008 hard flare. The last possibility could not be uncommon in a densely populated stellar system like a Globular cluster.
}
\\
As recently reported in literature by various authors (for a review see for example~\citet{Spruit} and reference therein) the state transitions in X-ray binaries { do not seem} to be governed by the mass accretion rate { only}. A second parameter, which nature is still unclear, seems to be involved in { the spectral} transitions.  This scenario is evident for BH binaries~\citep{homan01} that continuously shift the HID pattern, while for NS binaries this behaviour is less evident, even { though} the secular shifts { observed} in both ZETA and ATOLL sources (e.g. \cite{disalvo2003}) could be related to the { action of this unknown} parameter. { Therefore, one could speculate that also} the 2008 peculiar flare of \sor could be { an effect of} the action of the second parameter and not { directly related} to an increase of the mass accretion rate.

\section*{Acknowledgments}
We thank the anonymous referee for helpful comments that improved this manuscript. FC and MC  acknowledge financial contribution from the agreement ASI-INAF n.2017-14-H.O.

\bsp

\label{lastpage}

\end{document}